\documentclass[twocolumn,secnumarabic,amssymb, nobibnotes, aps, prd]{revtex4-2}

\usepackage{amsmath, braket, graphicx, xcolor, ulem}

\begin{document}

\title{Microwave Quantum Illumination with Optical Memory and\\Single-Mode Phase-Conjugate Receiver}%

\author{Sangwoo Jeon}%
\email[Sangwoo Jeon: ]{swjeon@add.re.kr}
\author{Jihwan Kim}
\author{Duk Y. Kim}
\author{Zaeill Kim}
\author{Taek Jeong}
\author{Su-Yong Lee}
\email[Su-Yong Lee: ]{suyong2@add.re.kr}
\affiliation{Agency for Defense Development, Daejeon 34186, Korea}

\begin{abstract}
Microwave quantum illumination with entangled pairs of microwave signal and optical idler modes, can achieve the sub-optimal performance with joint measurement of the signal and idler modes.
Here, we first propose a testbed of microwave quantum illumination with an optical memory which is simulated with a delay line in the idler mode.  
It provides how much an input two-mode squeezing is necessary to compensate the loss of the optical memory, while maintaining quantum advantage over coherent state.
When the memory is lossy, the input two-mode squeezing has to be higher through high cooperativity in the optical mode. 
Under the testbed, we propose a single-mode phase conjugate receiver that consists of a low-reflectivity beam splitter, an electro-optomechanical phase conjugator, and a photon number resolving detector. 
The performance of the newly proposed receiver approaches the maximum quantum advantage for local measurement. 
Furthermore, the quantum advantage is obtained even with an on-off detection while being robust against the loss of the memory.
\end{abstract}
\maketitle

\section{Introduction}
\label{Sec1}
Quantum illumination (QI) is a protocol that detects a low-reflectivity target using an entangled state within a fixed range, which outperforms classical illumination (CI) under conditions of high background noise and a low signal mean photon number \cite{Shapiro20,Sorelli21,Torrome24,Karsa23}.
Initially QI was proposed with single-photon-level entangled state and low background noise \cite{Lloyd08}, and then it was developed with gaussian states and high background noise \cite{Tan08}. Two-mode squeezed vacuum (TMSV) state presents a nearly optimal entangled state in QI \cite{DePalma18,Nair20,Bradshaw21}, whereas coherent state is theoretically a benchmark state.
QI with TMSV state can attain 6 dB quantum advantage over coherent state \cite{Tan08}, which requires collective measurement in a receiver \cite{Zhuang17,Shi22}.
Since the collective measurement has technical difficulties such as implementing feedforward process, beam splitter array, nonlinear interaction, memory, and so on, it is more feasible to consider local joint measurement on each pair of the returned signal mode and the idler mode. By local joint measurement, QI with TMSV state can achieve the sub-optimal 3 dB advantage \cite{Guha09,Sanz,Lee}.
Until now, several QI experiments with local measurement were implemented in microwave range \cite{Chang19,Luong20,Barzanjeh20,Assouly23} as well as in optical range \cite{Zhang15,England19,Aguilar19,Zhang20,Xu21}, where an optical (or a Josephson) parametric amplifier receiver could only obtain quantum advantage in experiment \cite{Zhang15,Assouly23}.

For the local joint measurement in QI, it is necessary to have a quantum memory to store the idler mode in order to be interfered with the reflected signal mode. There was a research from Reichert \textit{et al.} \cite{Reichert23} that quantum memory cannot be circumvented in the local joint measurement through a delay operation on a classical signal.
Thus, quantum memory is inevitable for QI.
The point is that the loss of the quantum memory in the idler mode highly affects the performance of QI.
Recently, Kim \textit{et al.} \cite{Kim23} demonstrated that the quantum memory loss of the idler mode dominates the performance of QI. 
When the memory loss is more than 50\%, QI cannot outperform CI under local joint measurement. 
Note that the upper bound of the memory loss is 75\% under collective measurement.
In optical range, the quantum memory is simulated with a delay line in the idler mode \cite{Lvovsky07}.
The current state-of-the-art efficiency of the optical delay line is about 0.14 dB/km, which therefore acts as a bottleneck on the realization of the long-ranged QI technology \cite{Tamura17}.

Although current technologies are more feasible in optical ranges, quantum advantage is more achievable in microwave range that is surrounded with high background noise. There were some trial experiments \cite{Hisatomi16,Regal18,Ihn20,Lauk20,Jeong23} to convert from microwave to optical wave and vice versa.
Alternatively, to take both benefits of optical and microwave ranges, Barzanjeh \textit{et al.} \cite{Barzanjeh15} proposed microwave QI by employing two electro-optomechanical (EOM) devices as a generation of input entangled microwave-optic fields at a transmitter and a phase-conjugated EOM system at a receiver. 
In this way, microwave QI can utilize the microwave signal and optical idler, having both benefits in achieving the quantum advantage: a large mean photon number of background noise in microwave signal frequency and the feasibility of optical components in optical idler frequency.

Since the originally proposed microwave QI assumed an ideal optical quantum memory \cite{Barzanjeh15}, it is important to consider a lossy optical memory in an idler mode, resulting in the minimum boundary of quantum advantage in microwave QI.
Here, we propose a testbed of microwave QI with a lossy optical memory. 
First, we observe the memory-loss effect on quantum advantage, providing an optimal input state to achieve the best quantum advantage regarding memory loss. Second, we propose a new receiver that is capable of operating in a lossy condition. 
We show that our receiver can approximately achieve the maximum quantum advantage for local joint measurement, even with implementing an on-off detection. 
Additionally, we show that the performance of the receiver is robust against the memory loss.

\begin{figure}[t]
    \centering
    \includegraphics[width=0.47\textwidth]{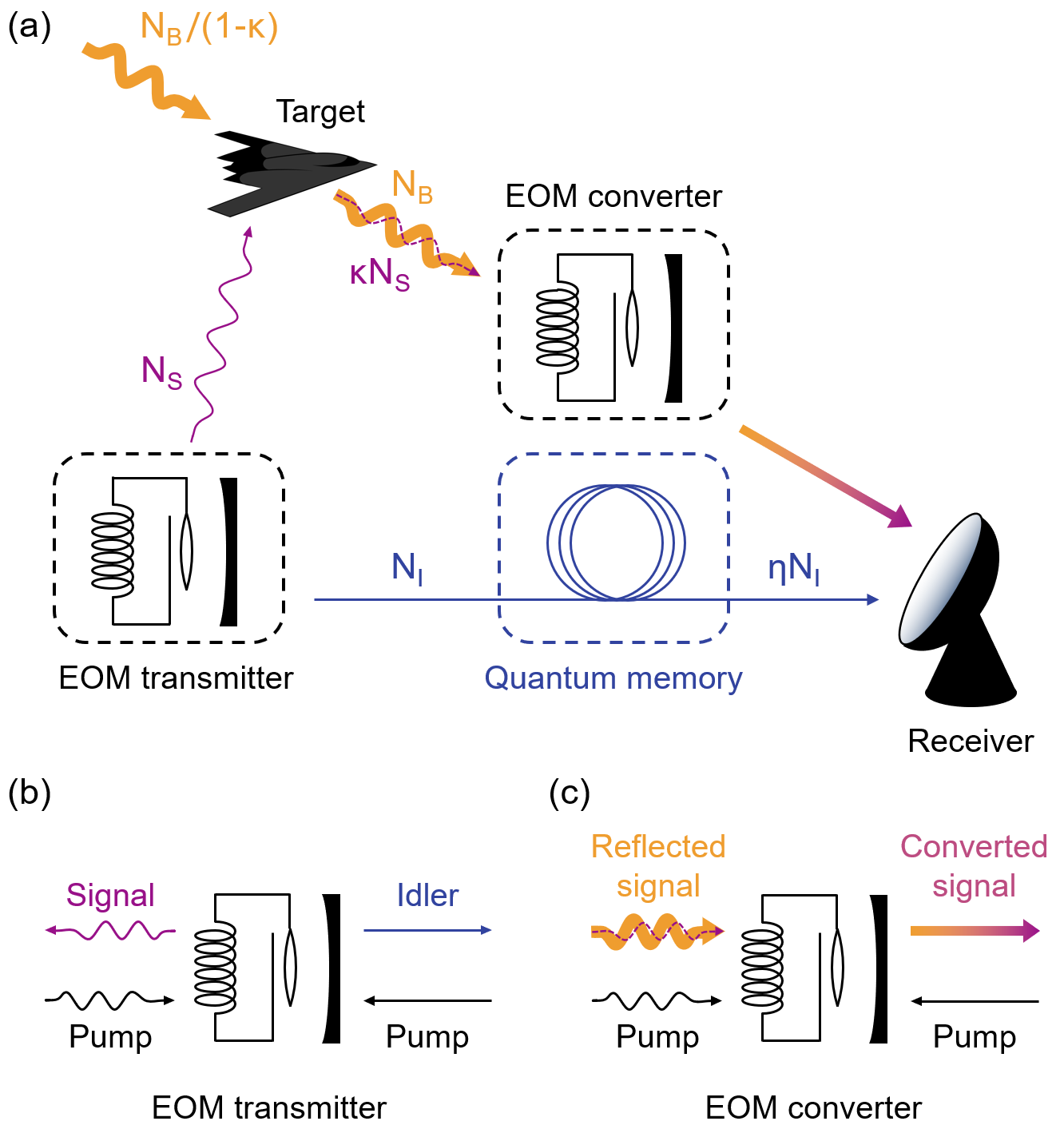}
    \caption{(a) Schematic of a microwave quantum illumination with quantum memory which is demonstrated as a delay line \cite{Lvovsky07}. (b) EOM transmitter produces entangled states, and (c) EOM converter transduces a reflected microwave signal into an optical signal. Curvy arrows denote the microwave modes, while straight arrows denote the optical modes.}
    \label{Fig1}
\end{figure}

Our paper is organized as follows.
Section \ref{Sec2} outlines our model of microwave QI with lossy optical memory under some conditions. 
Section \ref{Sec3} discusses the optimal conditions for the transmitter to attain the maximum quantum advantage for local joint measurement.
Section \ref{Sec4} introduces a novel receiver performing effectively despite memory loss and an on-off detection, along with an analysis of its performance. Section \ref{Sec5} draws the conclusion.

\section{Model}
\label{Sec2}

We consider microwave quantum illumination with quantum memory in Fig. \ref{Fig1}. (a), where an EOM transmitter generates entangled states that consist of pairs of microwave signal and optical idler.
EOM transmitter and converter consist of microwave and optical cavities which are coupled through a mechanical resonator, where the microwave and optical pumps are injected into the corresponding cavities as shown in Fig. \ref{Fig1}. (b) and (c).
In the absence of a target, the signal passes through the air and only a background thermal noise arrives at the EOM converter. 
When the target is present, the signal hits the target and arrives at the EOM converter along with the background thermal noise. 
The reflected signal carries the target information that is proportional to the target reflectivity $\kappa$. 
Upon the EOM converter, the reflected signal mode is frequency-converted to match the optical frequency of the idler. 
Meanwhile, the idler is sent to the quantum memory to preserve its state until joint measurement, where the quantum memory is represented by a delay line that is simply modeled as an effective transmission efficiency \cite{Ferraro}
$\eta=e^{-\gamma_I t}$ ($\gamma_I$: a dampinig rate, $t$: delay time).
Once both the converted signal mode and the stored idler mode are recombined, they are jointly measured in a receiver to detect a target. We assume that the receiver performs the local joint measurement.

In Fig. \ref{Fig1}, 
a target is modeled as a beam splitter with a reflectivity $\kappa$, so that the transformation by the target interaction is given by 
$\hat{b}_S=\sqrt{\kappa}\hat{a}_S+\sqrt{1-\kappa}\hat{a}_B$, where $\hat{a}_{S(I)}$, $\hat{a}_B$, and $\hat{b}_S$ are the annihilation operators of signal(idler), background thermal noise, and returned signal modes, respectively.
Since the thermal noise is independent of the target reflectivity after interacting with the target, the mean photon number of the thermal noise is defined as $N_B/(1-\kappa)$ and then the returned signal mode presents the mean photon number of $\langle\hat{b}_S^\dagger\hat{b}_S\rangle=\kappa N_S+N_B$. 
The best quantum advantage is obtained in the range of $\kappa,N_S\ll 1\ll N_B$, which validates microwave signal under high background thermal noise. Note that it is negligible for thermal noise in the memory since the idler is in optical regime. 

QI system discriminates whether $\kappa>0$ (target presence) or $\kappa=0$ (target absence) by measuring an observable $\hat{O}$ on the joint returned signal and idler modes. The performance is described with a detection error probability.
We focus on verifying the target presence by comparing mean of $M$ observables $\{O^{(k)}:1\leq k\leq M\}$ with a decision threshold $R_{th}$, which decides $\kappa>0$ if $\bar{O}:=\frac{1}{M}\sum_k O^{(k)}>R_{th}$ and $\kappa=0$ if $\bar{O}\leq R_{th}$. Then, the detection error probability $P_{err}$ with $\hat{O}$ is given by
\begin{equation}
	P_{err}(\hat{O}):=\min_{R_{th}}[Pr(\bar{O}>R_{th}|\kappa=0)+Pr(\bar{O}\leq R_{th}|\kappa>0)],
\end{equation}
which decreases exponentially with the number of modes $M$, so that $P_{err}(\hat{O})\sim\exp(-{\rm SNR(\hat{O})}M)$ \cite{Chernoff52}, where the signal-to-ratio (SNR) is defined as
\begin{equation}
    {\rm SNR}(\hat{O}):=\frac{(\langle\hat{O}\rangle_{\kappa>0}-\langle\hat{O}\rangle_{\kappa=0})^2}{2\left( \sqrt{\langle\Delta\hat{O}^2\rangle_{\kappa>0}}+\sqrt{\langle\Delta\hat{O}^2\rangle_{\kappa=0}}\right)^2}.
    \label{SNR}
\end{equation}
Under local measurement with the observable, the maximum performance of the QI is defined as the maximum error exponent $\gamma_{\rm QI}:=\max_{\hat{O}} {\rm SNR}(\hat{O})$.

Although a TMSV state is a nearly optimal input state for QI, 
the microwave QI of Ref. \cite{Barzanjeh15} generates a two-mode squeezed thermal state that is respresented as
\begin{equation}
    \rho=\hat{S}_{SI}(r)[\rho_{th}(\nu_S)\otimes\rho_{th}(\nu_I)]\hat{S}_{SI}^\dagger(r),
\end{equation}
where $\hat{S}_{SI}(r)=\exp[r(\hat{a}_S\hat{a}_I-\hat{a}_S^\dagger\hat{a}_I^\dagger)]$ and $r$ is a squeezing parameter of the two-mode squeezing operator.
$\rho_{th}(\nu)$ is a thermal state with a symplectic eigenvalue $\nu$, where $\nu_S$ and $\nu_I$ represent the symplectic eigenvalues of the signal and idler modes. Since the symplectic eigenvalues are the eigenvalues of the symplectic decomposition of the covariance matrix of the joint signal-idler state, it is given by $\nu_i=2N_{th,i}+1$ for $i\in\{S,I\}$, where $N_{th, i}$ is the mean photon number of $\rho_{th}(\nu_i)$. This implies that a two-mode squeezed thermal state with high symplectic eigenvalues is generated by squeezing a thermal state with high temperature,  whereas the TMSV state gives the minimum symplectic eigenvalue $\nu_S=\nu_I=1$. 
In the same scenario, it is reasonable to consider a displaced thermal state as a classical benchmark for microwave QI \cite{Karsa21}, but here we take a strict benchmark with a coherent state.

\section{Optimal input state for quantum advantage}
\label{Sec3}
\begin{figure}[t]
    \centering
    \includegraphics[width=0.48\textwidth]{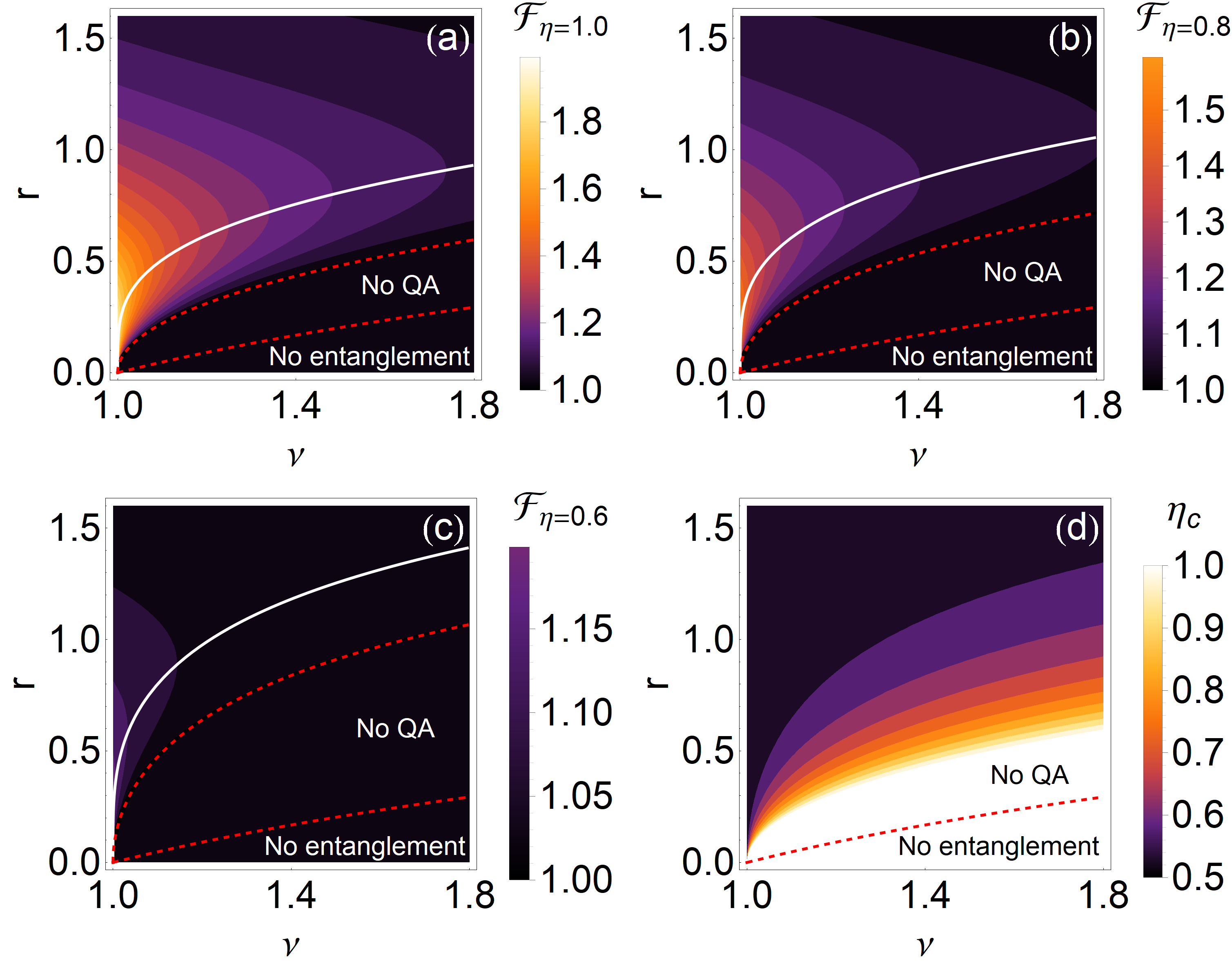}
    \caption{Quantum advantage ($\mathcal{F}$) and the critical memory efficiency ($\eta_c$) of QI with a two-mode squeezed thermal state as a function of $\nu$ and $r$ at $\kappa=0.01$ and $N_B=600$. A set of three figures (a), (b) and (c) shows the quantum advantage figure of QI with memory efficiences of $\eta=1$, 0.8, and 0.6. The solid white line shows the optimal squeezing parameter. The upper dashed red line shows a boundary of $\mathcal{F}=1$, so that the region below shows no quantum advantage (QA). (d) shows the critical memory efficiency. The lower dashed red line shows a boundary of entanglement.}
    \label{Fig2}
\end{figure}
In order to evaluate the quantum advantage, we employ the maximum error exponent $\gamma_{\rm QI}$ for QI with an optimal input state that is compared with $\gamma_{\rm CI}$ for CI with a coherent state. Under the same signal mean photon number, let $\mathcal{F}=\gamma_{\rm QI}/\gamma_{\rm CI}$ be a figure of quantum advantage. 
Then $\mathcal{F}>1$ implies the quantum advantage. At $N_S\ll1\ll N_B$, the quantum advantage can approach $\mathcal{F}= 2$ with local measurement, which represents the sub-optimal 3 dB quantum advantage.

In this section, we present the optimal condition of an input state to achieve the best quantum advantage under memory loss. 
The input state has two different parameter domains as follows: First, the input state is a function of symplectic eigenvalue and two-mode squeezing parameter. Second, the input state is a function of cooperativities, where it is generated from the EOM transmitter.

\subsection{Input two-mode squeezed thermal state}

\begin{figure}[t]
    \centering
    \includegraphics[width=0.33\textwidth]{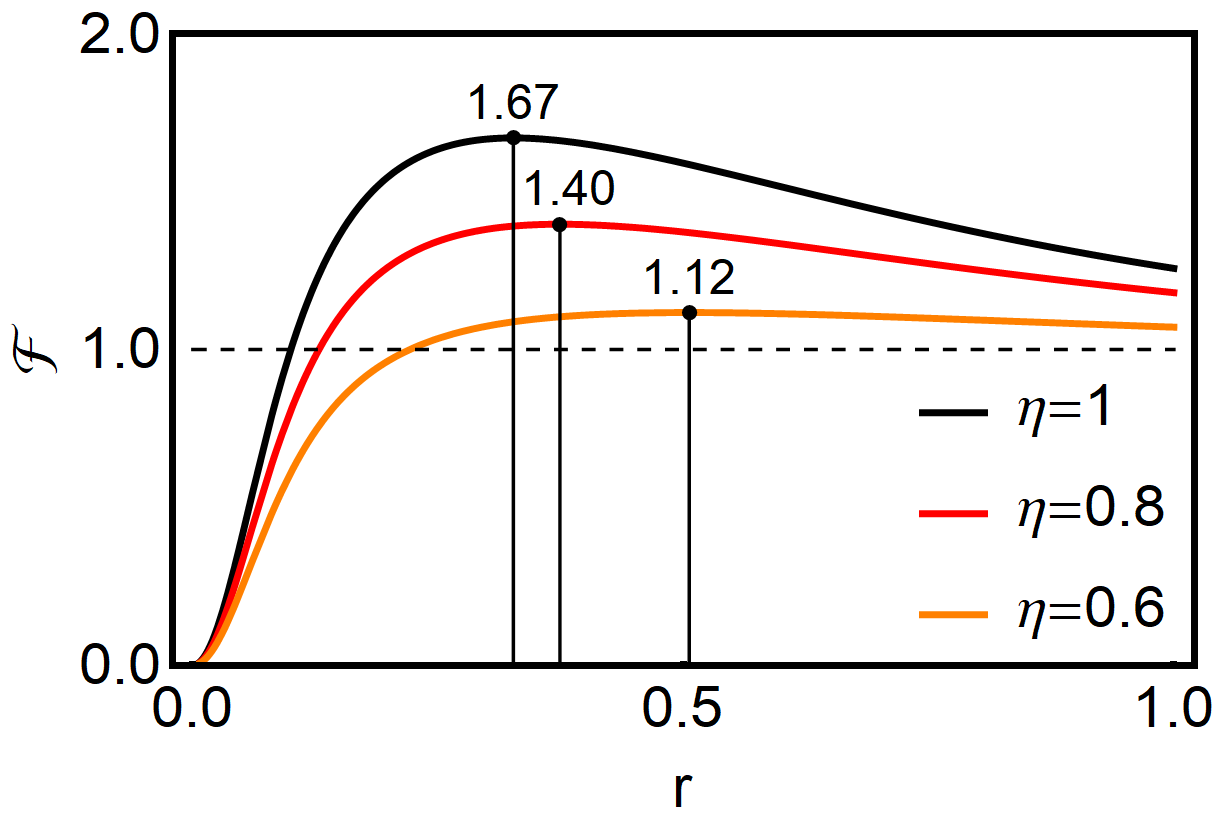}
    \caption{Quantum advantage ($\mathcal{F}$) as a function of $r$ at $\nu=1.02$ (two-mode squeezed thermal), $\kappa=0.01$, and $N_B=600$ with $\eta=$1 (black), 0.8 (red), and 0.6 (orange). Each vertical line indicates the optimal squeezing at corresponding $\eta$, where the maximum $\mathcal{F}=1.67$, 1.40, and 1.12 are achieved at $r=0.327$, 0.374, and 0.506 respectively.}
    \label{Fig3}
\end{figure}

We show that an input two-mode squeezing parameter has to be increased to compensate the loss from optical memory, where explicit formulas are given in Appendix \ref{App1}. We describe the maximum error exponents of QI and CI as follows,
\begin{align}
\gamma_{\rm QI}&=\frac{\eta \kappa N_{S,I}^2}{2(1+\eta N_I+N_B+2\eta N_I N_B)},\\
\gamma_{\rm CI}&=\frac{\kappa N_S}{4N_B+2},
\end{align}
where $N_{S,I}:=\langle\hat{a}_S\hat{a}_I+\hat{a}_I\hat{a}_S\rangle/2$ \cite{Kim23}. For the convenience of visualization, we assume the symplectic eigenvalues of the signal and idler modes are the same, \textit{i.e.}, $N_S=N_I$, noting that the result is similar to the unbalanced case of $N_S\neq N_I$. Then, by substituting $N_S=\frac{1}{2}(\nu\cosh2r-1)$ and $N_{S,I}=\frac{1}{2}\nu\sinh2r$, $\mathcal{F}$ is plotted as a function of $\nu$ and $r$ in Fig. \ref{Fig2}. (a)-(c). Each figure shows $\mathcal{F}$ with different memory efficiencies of 1, 0.8, and 0.6. The region above the upper dashed line ($\mathcal{F}>1$) shows the existence of quantum advantage, whereas the region below ($\mathcal{F}\leq 1$) represents no quantum advantage. The lower dashed line shows a boundary of the entanglement, where the region above the line represents the existence of entanglement. This shows that entanglement does not ensure the quantum advantage. As the squeezing parameter increases, $\mathcal{F}$ increases until it intersects the optimal squeezing line, as shown with the solid white line. Beyond the point, $\mathcal{F}$ gets smaller while maintaining the advantage. 

We address some notable phenomena under memory loss. First, memory loss lowers the maximum value of quantum advantage that is  given by $2\eta$ ($\eta$: memory efficiency), where the maximum value is achieved at $\nu=1$ and $r\to 0$. Second, memory loss increases the optimal value of the input two-mode squeezing parameter. Third, memory loss increases the area of no quantum advantage. 
The results are clearly demonstrated in Fig. \ref{Fig3}, where each curve represents the quantum advantage as a function of $r$ at $\nu=1.02$, corresponding to a cross section of Fig. \ref{Fig2}. (a)-(c).

To further observe the relation between the quantum advantage and the memory loss, we examine a critical memory efficiency $\eta_c$ that is a function of an input state, referring to the minimum memory efficiency required to achieve the quantum advantage. By solving $\mathcal{F}=1$ as an equation of $\eta$, $\eta_c$ can be approximately represented by
\begin{equation}
\label{eta_c}
\eta\simeq\frac{n_S}{2(n_{S,I}^2-n_S n_I)}=:\eta_c.
\end{equation}
Fig. \ref{Fig2}. (d) shows $\eta_c$ in the domain of $\nu$ and $r$, substituting the mean photon number terms. The white colored region shows where the quantum advantage is unachievable even with the ideal memory. Under fixed $\nu$, $\eta_c$ is a decreasing function of $r$ with a lower bound of 1/2. This implies that a highly squeezed two-mode state is resilient to the memory loss. 
Note that the lower bound of 1/2 in the memory efficiency implies the maximum detection range being 21.21 km while keeping the quantum advantage, which is achievable by utilizing an optical delay line at a 1560 nm wavelength \cite{Tamura17}.

\begin{figure}
    \centering
    \includegraphics[width=0.48\textwidth]{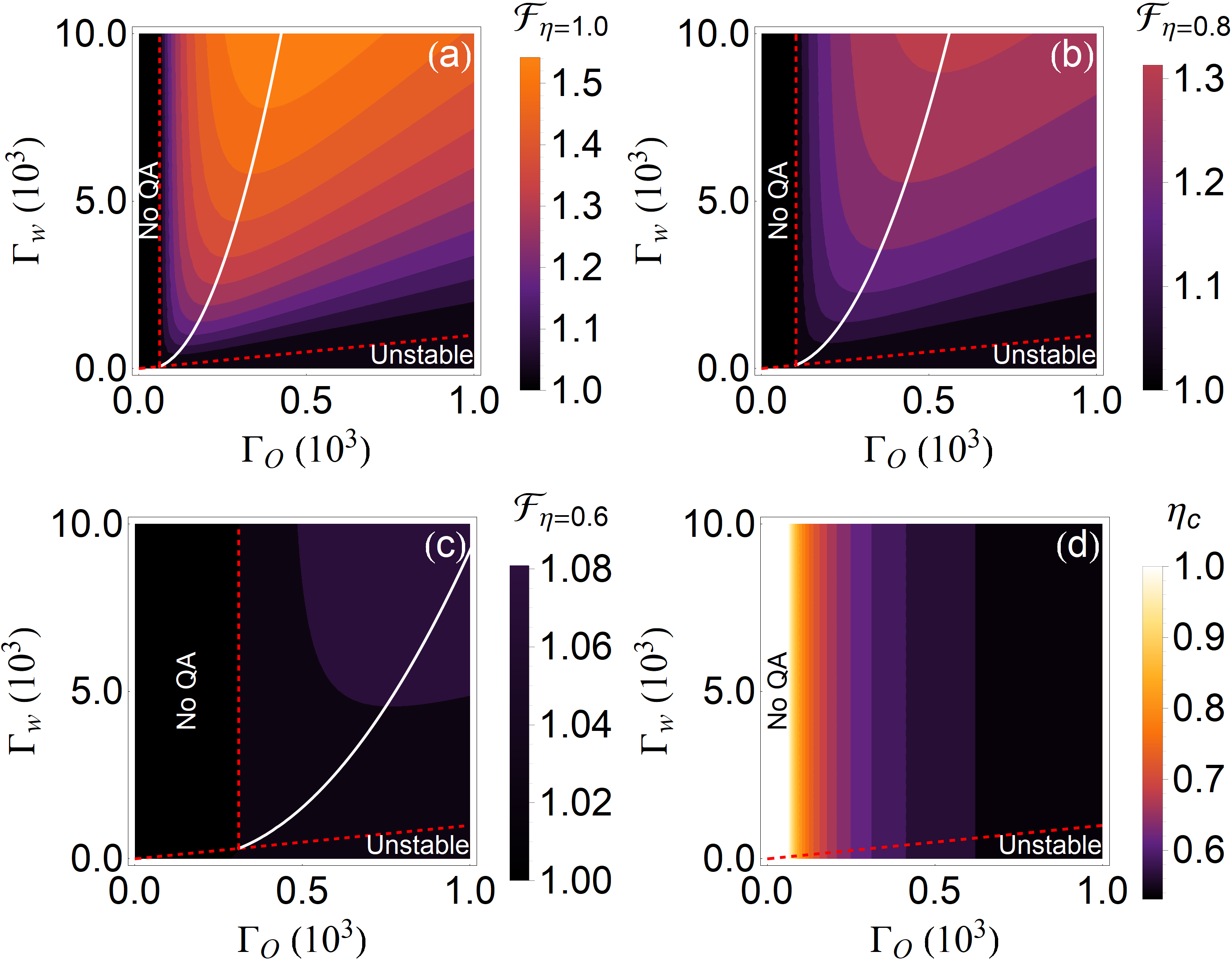}
    \caption{Quantum advantage ($\mathcal{F}$) and the critical memory efficiency ($\eta_c$) of QI with an input state from EOM transmitter at $\kappa=0.01$ and $N_B=600$. A set of three figures (a), (b) and (c) show the quantum advantage figure of QI with memory efficiences of $\eta=1$, 0.8, and 0.6. The solid white line shows the optimal $\Gamma_o$. The vertical dashed red line shows a boundary of $\mathcal{F}=1$. The lower dashed red line shows the boundary of stable region.
(d) shows the critical memory efficiency.  Experimentally achievable parameters are employed as in previous researches \cite{Barzanjeh15,Ebrahimi22}, where temperature of EOM devices with $T_{\rm EOM}=30~{\rm mK}$, frequency of mechanical and microwave of $\omega_M/2\pi=10~{\rm MHz}$ and $\omega_w/2\pi=10~{\rm GHz}$, and the wavelength of driving optical laser with 1064 nm. }
    \label{Fig4}
\end{figure}

\subsection{EOM-generated input state}

\begin{figure}
    \centering
    \includegraphics[width=0.33\textwidth]{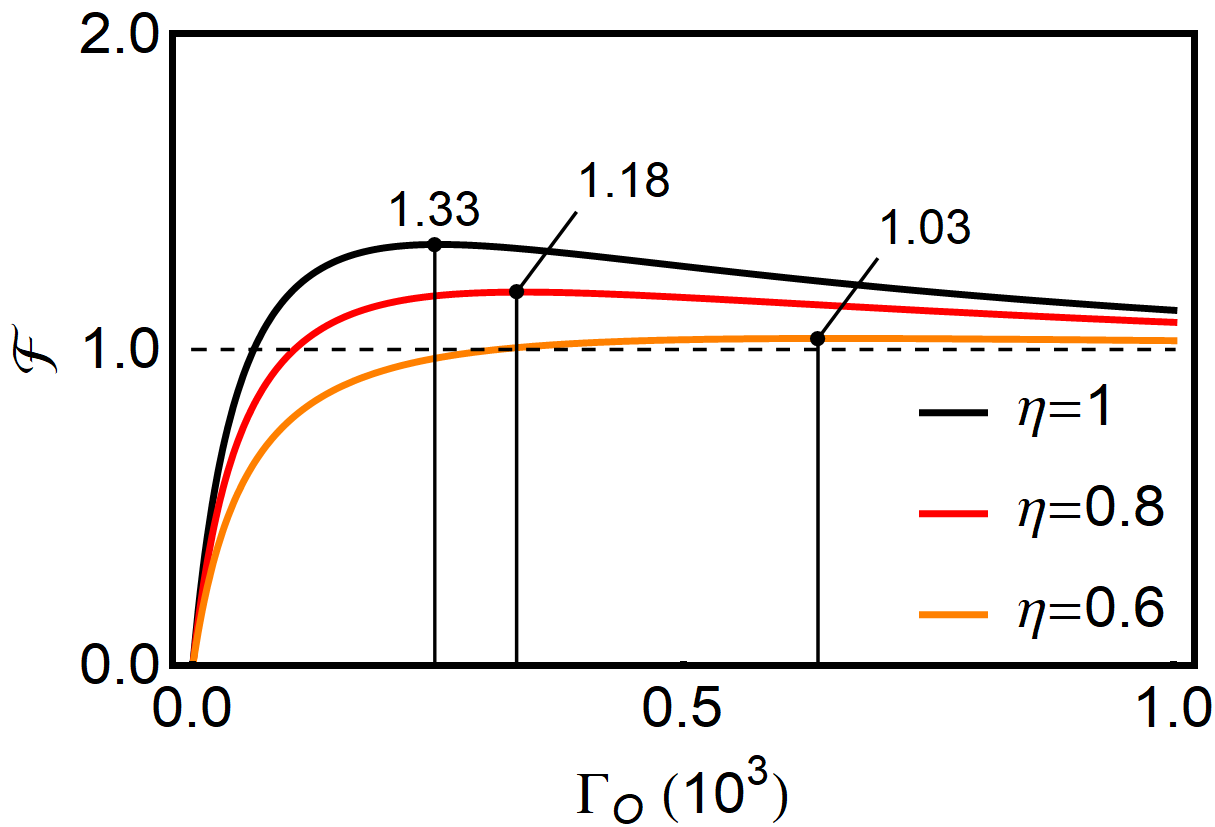}
    \caption{Quantum advantage ($\mathcal{F}$) as a function of $\Gamma_o$ at $\Gamma_w=3000$, $\kappa=0.01$, and $N_B=600$ with $\eta=$1 (black), 0.8 (red), and 0.6 (orange). Each vertical line indicates the optimal $\Gamma_o$ at corresponding $\eta$, where the maximum $\mathcal{F}=1.33$, 1.18, and 1.03 are achieved at $\Gamma_o=247$, 330, and 637 respectively.}
    \label{Fig5}
\end{figure}

We now consider the practical Gaussian input state generated by an EOM device, where the state is determined by cooperativities of the optical and microwave cavities, namely $\Gamma_o$ and $\Gamma_w$. 
Here, the cooperativities are defined as $\Gamma_j=G_j^2/\kappa_j\gamma_M$ for $j\in\{o,w\}$, where $G_j$ is a coupling rate, while $\kappa_j$ and $\gamma_M$ are dissipation rates of the corresponding cavities and resonator \cite{Barzanjeh15}.
From the relations between mean photon number and cooperativities, we can illustrate $\mathcal{F}$ as a function of $\Gamma_o$ and $\Gamma_w$ in Fig. \ref{Fig4}. (a)-(c).  

The memory loss reduces the maximum value and the area of quantum advantage, such that it is required to increase $\Gamma_o$ to achieve the quantum advantage with increasing memory loss, as shown in Fig. \ref{Fig5}.
It emphasizes the importance of a lossless quantum memory, since high cooperativities are challenging to achieve.
In Fig. \ref{Fig4}. (d), we show that the critical  memory efficiency $\eta_c$ solely depends on $\Gamma_o$, where
\begin{equation}
\label{eta_c_coop}
\eta_c=\frac{\Gamma_o+N_M^T}{2\Gamma_o}
\end{equation}
and $N_M^T$ is the mean photon number of a thermal state with frequency and temperature of the mechanical resonator.
Since the boundary of quantum advantage is given by the solution of $\eta_c=\eta$, the minimum value of $\Gamma_o=N_M^T/(2\eta-1)$ to achieve the quantum advantage is proportional to $N_M^T$.
This implies that the low-temperature environment of the EOM transmitter is crucial. Additional information and formulas are given in Appendix \ref{App2}.

\section{Single mode phase-conjugate receiver}
\label{Sec4}
Any QI setup does not always achieve the maximum quantum advantage, such that it is necessary to investigate the performance of QI with a practical receiver. Here, we employ $\mathcal{F}_{\rm receiver}={\rm SNR}(\hat{O}_{\rm receiver})/\gamma_{\rm CI}$ as the figure of quantum advantage, with an observable of the receiver $\hat{O}_{\rm receiver}$. 

\begin{figure}
    \centering
    \includegraphics[width=0.48\textwidth]{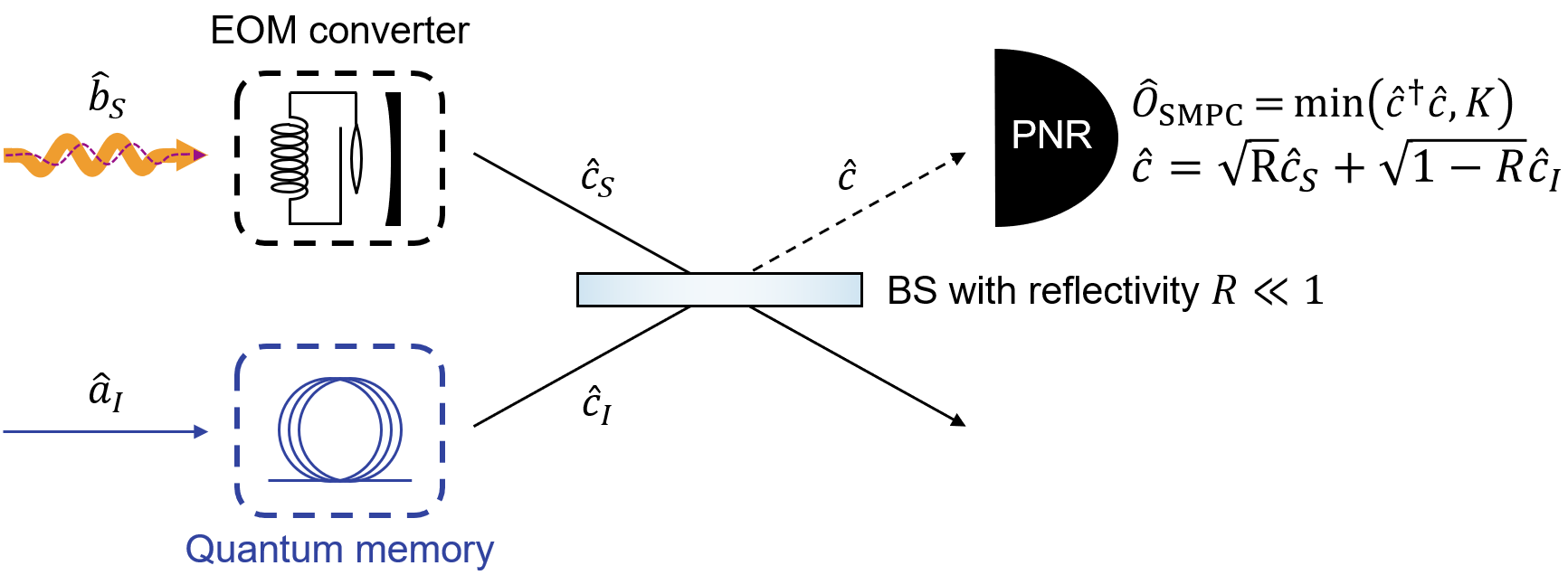}
    \caption{Schematic of the SMPC receiver. Frequency-converted signal mode and stored idler gets interfered by a low-reflectivity beam splitter and yields an output mode arriving at the PNR receiver.}
    \label{Fig6}
\end{figure}

In the microwave QI of Ref. \cite{Barzanjeh15}, a phase-conjugate (PC) receiver has been suggested by addressing the EOM device as a frequency-converting phase conjugator. PC receiver interferes the frequency-converted signal mode and stored idler mode by a 50:50 beam splitter and measures the photon number difference of the output modes with photon number detectors. Theoretically, the PC receiver achieves the quantum advantage $\mathcal{F}_{\rm PC}$ up to $3$ dB \cite{Guha09}. However, experimentally, it has not been completely realized while digitally being achieved with no quantum advantage \cite{Barzanjeh20}. The obstacle to implementing the PC receiver comes from the necessity of a photon number resolving (PNR) detector with high resolution, since the PC receiver setup assumes to fully count the frequency converted background thermal noise. 
Actually, it is a challenge to implement a high-resolution PNR detector. 

We propose a single-mode PC (SMPC) receiver that consists of an EOM device, a low-reflectivity beam splitter, and a finite-resolution PNR detector, as shown in Fig. \ref{Fig6}. The reflected signal mode $\hat{b}_S$ goes through the EOM converter, resulting in its frequency conversion into the optical frequency. The frequency-converted signal mode $\hat{c}_{S}$ and the stored idler mode $\hat{c}_{I}$ are interfered at the beam splitter with low reflectivity $R\ll 1$. Given the output mode $\hat{c}=\sqrt{R}\hat{c}_S+\sqrt{1-R}\hat{c}_I$, the PNR detector measures the photon number of the mode up to its resolution $K$. The corresponding observable of the PNR detector is given by $\hat{O}_{\rm SMPC}=\min(\hat{c}^\dagger\hat{c},K)$. 

\begin{figure}[t]
    \centering
    \includegraphics[width=0.48\textwidth]{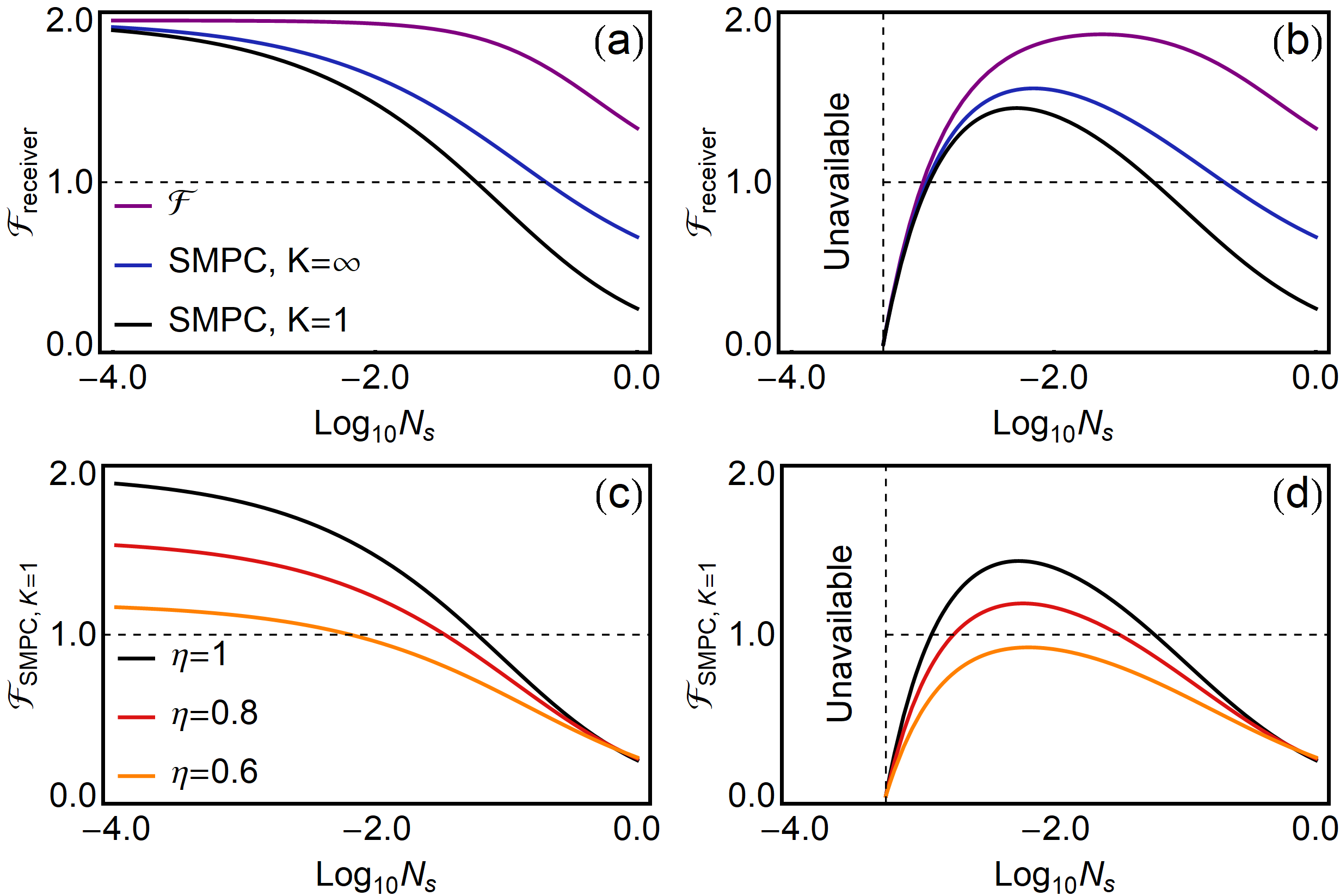}
    \caption{Performance of SMPC receiver as a function of log-scaled $N_S$. Upper two figures with (a) $\nu=1$ (TMSV) and (b) $\nu=1.001$ (two-mode squeezed thermal) show the quantum advantage corresponding to the maximum error exponent $\mathcal{F}$ (purple), SMPC with $K=\infty$ (blue) and 1 (black), assuming $\eta=1$. Lower two figures with (c) $\nu=1$ and (d) $\nu=1.001$ show the quantum advantage by taking the SMPC reciever with K=1, $\mathcal{F}_{{\rm SMPC}, K=1}$, at different memory efficiences of $\eta=$1 (black), 0.8 (red) and 0.6 (orange). We assume the same environment from Fig. \ref{Fig4}. Cooperativities of the EOM converter are given by $\Gamma_o=60$ and $\Gamma_w=600$.}
    \label{Fig7}
\end{figure}

By choosing appropriate cooperativities of the EOM converter and the reflectivity $R=\sqrt{N_{c, I}}/N_{c, S}$ with $N_{c,i}:=\langle\hat{c}_{i}^\dagger\hat{c}_{i}\rangle$ for $i\in\{S, I\}$, the SMPC receiver can fully achieve 3 dB quantum advantage, \textit{i.e.} $\mathcal{F}_{\rm SMPC}= 2$. Note that $N_{c,S}$ is assumed to be invariant whether $\kappa=0$ or $\kappa>0$. We emphasize that the SMPC receiver can achieve the possible maximum quantum advantage regardless of $K$, which is proved in Appendix \ref{App3}. 
Suppose the quantum-advantage regime of $\Gamma_o,\Gamma_w\gg 1$, which validates to assume $\nu_S=\nu_I=\nu$ from $N_S\simeq N_I$.
Fig. \ref{Fig7}. (a) shows $\mathcal{F}_{\rm SMPC}$ corresponding to $K=1$ and $\infty$ assuming the input state as a TMSV state. 
As shown in the figure, even in the case of $K=1$, \textit{i.e.} when employing the on-off photon detection, the SMPC receiver achieves the 3 dB advantage in a low-$N_S$ regime. 
 When the input signal includes the internal thermal noise, \textit{i.e.} $\nu>1$, the receiver may achieve the lower quantum advantage as shown in Fig. \ref{Fig7}. (b). Note that when $\nu>1$, the signal photon number is lower-bounded to $(\nu-1)/2$, so that the unavailable region exists in the domain.

In Fig. \ref{Fig7}. (c) and (d), we show that the SMPC receiver is robust against memory loss. 
It exhibits the quantum advantage by $\mathcal{F}_{{\rm SMPC},K=1}$ in different memory efficiencies, where the TMSV state achieves the maximally feasible quantum advantage of $2\eta$,
implying that the SMPC receiver works well regardless of memory loss.

\section{Conclusion}
\label{Sec5}

We presented a testbed of microwave QI with a lossy optical memory. To take quantum advantage under the memory loss, 
it is necessary to increase the input two-mode squeezing parameter that is proportional to increase the cooperativity of an optical cavity in an EOM trasmitter. It was quantitatively analyzed by maximum error exponents for local measurement.
Under the testbed, we proposed an SMPC receiver that can achieve the maximum quantum advantage for local measurement. 
Our new receiver takes quantum advantage even with an on-off detection, which is also robust against the optical memory loss.


Since the memory loss profoundly affects the quantum advantage, our contribution lies on enhancing the performance of microwave QI with quantum  memory in a practical setup. 
We expect that our idea can be further developed by taking a feedback method that was considered in microwave QI \cite{Ebrahimi22}.

\begin{acknowledgments}
This work was supported by a grant to Defense-Specialized Project funded by Defense Acquisition Program Administration and Agency for Defense Development.
\end{acknowledgments}

\appendix
\section{Quantum advantage in two-mode squeezed thermal state input}
\label{App1}
We provide equations on quantum advantage and optimal parameters of QI with two-mode squeezed thermal state input. 
Under $\kappa\ll 1$ and $N_S\ll N_B$, $\mathcal{F}=\gamma_{\rm QI}/\gamma_{\rm CI}$ can be approximated as follows.
\begin{equation}
	\mathcal{F}\simeq\frac{2\eta N_{S,I}^2}{N_S(1+2\eta N_I)}
	\label{F}
\end{equation}
By substituting $N_S=N_I=\frac{1}{2}(\nu\cosh2r-1)$ and $N_{S,I}=\frac{1}{2}\nu\sinh2r$,
\begin{equation}
	\mathcal{F}=\frac{\eta\nu^2\sinh^2 2r}{(\nu \cosh 2r-1)(\eta\nu\cosh 2r-\eta +1)}.
\end{equation}
The boundary of the quantum advantage is solved at $\mathcal{F}=1$, which is given by
\begin{equation}
r_{\rm QA}(\nu)=\frac{1}{2}\cosh^{-1}\left[\frac{\eta\nu^2+\eta-1}{(2\eta-1)\nu}\right].
\end{equation}
The optimal squeezing parameter can be given by solving $\frac{\partial \mathcal{F}}{\partial r}=0$. 
We obtain
\begin{align}
r_{\rm optimal}(\nu)&=\nonumber
\\ \frac{1}{2}\cosh^{-1}&\left[\frac{\eta\nu^2+\eta-1+\sqrt{(\nu^2-1)((\nu^2-1)\eta^2+2\eta-1)}}{(2\eta-1)\nu}\right].
\end{align}
It can be seen that $r_{\rm QA}(\nu)$ and $r_{\rm optimal}(\nu)$ are both decreasing functions of $\eta$ for $\eta\in(\frac{1}{2}, 1]$ and $\nu>1$. 
Thus, high $r$ leads to low $\eta_c$. 
By substituting $N_S$ and $N_{S,I}$ in Eq. \ref{eta_c}, we can also write down $\eta_c$ as follows,
\begin{equation}
\eta_c=\frac{1}{2-\frac{\nu^2-1}{\nu\cosh 2r -1}}.
\end{equation}
Since $\eta_c$ is a decreasing function of $r$, the highly-squeezed states are resilient against the memory loss.

\section{Quantum advantage in microwave QI}
\label{App2}
We briefly introduce the microwave QI setup with reviewing Ref. \cite{Barzanjeh15} and provide equations on the quantum advantage of the setup with considering the memory loss. 
Let $\omega_w$, $\omega_o$, and $\omega_M$ be the resonance frequencies of microwave cavity, optical cavity, and mechanical resonator, respectively. 
By injecting driving pumps to the microwave and optical cavities with each detuned resonance frequency respectively, the output modes from the EOM system are given by
\begin{align}
\hat{a}_S&=A_w\hat{d}_{w,{\rm in}}-B\hat{d}_{o,{\rm in}}^\dagger-C_w\hat{d}_{M,{\rm in}},\\
\hat{a}_I&=B\hat{d}_{w,{\rm in}}^\dagger+A_o\hat{d}_{o, {\rm in}}-C_o\hat{d}_{M,{\rm in}}^\dagger,
\end{align}
where the effective detuning on the microwave and optical cavities are $\Delta_w=\omega_M$ and $\Delta_o=-\omega_M$, respectively \cite{Barzanjeh11}. Here, $\hat{d}_{i, {\rm in}}$ indicates the internal thermal mode of a cavity with temperature $T_{\rm EOM}$ and frequency $\omega_{i}/2\pi$, such that $N_{i}^T:=\langle\hat{d}_{i, {\rm in}}^\dagger\hat{d}_{i, {\rm in}}\rangle=[\exp(\hbar\omega_i/(k_B T_{\rm EOM}))-1]^{-1}$ for $i\in\{w, o, M\}$, while the coefficients
\begin{align}
A_w&=\frac{1-(\Gamma_w+\Gamma_o)}{1+\Gamma_w-\Gamma_o}\\
A_o&=\frac{1+(\Gamma_w+\Gamma_o)}{1+\Gamma_w-\Gamma_o}\\
B&=\frac{2\sqrt{\Gamma_w \Gamma_o}}{1+\Gamma_w-\Gamma_o}\\
C_o&=\frac{2i\sqrt{\Gamma_o}}{1+\Gamma_w-\Gamma_o}\\
C_w&=\frac{2i\sqrt{\Gamma_w}}{1+\Gamma_w-\Gamma_o}
\end{align}
are given as a function of the cooperativities $\Gamma_w$ and $\Gamma_o$. Since we suppose the EOM in a dilution fridge with its temperature of mK order, the mean photon number of thermal-noise is extremely lower than $1$, such as $N_w^T, N_o^T\ll 1$. Thus, the mean photon number and correlation term of signal and idler are given as follows.
\begin{align}
N_S&=B^2+|C_w|^2N_M^T\\
N_I&=B^2+|C_o|^2(N_M^T+1)\\
N_{S,I}&=A_wB+C_wC_o(N_M^T+1)
\end{align}
From Eq. (\ref{F}), we obtain
\begin{equation}
\mathcal{F}=\frac{2\eta\Gamma_o(1+\Gamma_o+\Gamma_w+2N_M^T)^2}{(\Gamma_o+N_M^T)[(1-\Gamma_o+\Gamma_w)^2+8\eta\Gamma_o(1+\Gamma_w)+8\eta\Gamma_o N_M^T]}
\end{equation}
with assuming $\Gamma_w>\Gamma_o$. By solving $\frac{\partial\mathcal{F}}{\partial\Gamma_w}=0$ and $\mathcal{F}=1$, we have the optimal $\Gamma_o$ and critical memory efficiency
\begin{align}
\Gamma_{o,{\rm optimal}}&=\frac{1}{2}\left(\frac{N_M^T}{2\eta-1}+\sqrt{\frac{N_M^T\Gamma_w}{2\eta-1}}\right),\\
\eta_c&=\frac{\Gamma_o+N_M^T}{2\Gamma_o}.
\end{align}

\section{Quantum advantage of SMPC receiver}
\label{App3}
We show that the SMPC receiver can achieve the 3 dB quantum advantage by employing a low-reflectivity beam splitter, finite-resolution PNR receiver, and TMSV input state with $N_S=N_I\ll 1$. We first show that with appropriate $\Gamma_o$ and $\Gamma_w$, $\sqrt{N_{c,I}}\ll N_{c,S}$ can be satisfied, which leads to $R=\sqrt{N_{c,I}}/N_{c,S}\ll 1$. The input modes of the beam splitter are
\begin{align}
\hat{c}_S&=B\hat{b}_S^\dagger+A_o\hat{d}_{o,{\rm in}}'-C_o\hat{d}_{M,{\rm in}}'^\dagger,\\
\hat{c}_I&=\sqrt{\eta}\hat{a}_I+\sqrt{1-\eta}\hat{a}_V,
\end{align}
where $\hat{a}_V$ is a vacuum mode. The mean photon number and correlation term of the modes are given by
\begin{align}
N_{c,S}&=B^2(N_B+1)+|C_o|^2(N_M^T+1),\\
N_{c,I}&=\eta N_I=\eta N_S,\\
N_{c,S,I}:&=\frac{1}{2}\langle\hat{c}_S^\dagger\hat{c}_I+\hat{c}_I^\dagger\hat{c}_S\rangle\nonumber
\\&=B\sqrt{\eta\kappa}N_{S,I}=B\sqrt{\eta\kappa N_S(N_S+1)}.
\end{align}
Under the condition of $\Gamma_o, \Gamma_w\gg 1$, $|C_o|^2\ll B^2$ holds. Since we suppose the EOM in a dilution fridge, the mean photon number of the mechanical cavity thermal mode satisfies $N_M^T<N_B$ ($N_M^T\simeq 62$ in 30 mK), which leads to $N_{c,S}\simeq B^2N_B$. Thus, $\sqrt{N_{c,I}}\ll N_{c,S}$ is equivalent to $\sqrt{N_{c,I}}/N_B\ll B^2$. The corresponding cooperativities that yield $B=\Omega(N_B^{-1/2})$ satisfy the condition. Note that our choice of $\Gamma_o=60$ and $\Gamma_w=600$ used on the numerical calculation suffices the condition by $B>N_B^{-1/2}$.

Now, we show that the receiver can achieve the sub-optimal 3 dB quantum advantage. The output mode arriving in the detector is given as follows.
\begin{equation}
\hat{c}=\sqrt{R}\hat{c}_S+\sqrt{1-R}\hat{c}_I.
\end{equation}
Thus, the mean photon number of the output mode is given by
\begin{align}
N_{c,\kappa=0}:&=\langle\hat{c}^\dagger\hat{c}\rangle_{\kappa=0}=RN_{c,S}+(1-R)N_{c,I},\\
N_{c,\kappa>0}:&=\langle\hat{c}^\dagger\hat{c}\rangle_{\kappa>0}=\langle\hat{c}^\dagger\hat{c}\rangle_{\kappa=0}+2\sqrt{R(1-R)}N_{c,S,I}.
\end{align}
Note that $N_{c,\kappa=0}\simeq N_{c,\kappa>0}\ll1$ holds under the given $R$.
Recalling Eq. (\ref{SNR}), our goal is to get mean and variance of the observable $\hat{O}_{\rm SMPC}=\min(\hat{c}^\dagger\hat{c},K)$, under the conditions of $\kappa=0$ and $\kappa>0$. Denoting $q_c=\langle\hat{c}^\dagger\hat{c}\rangle/(\langle\hat{c}^\dagger\hat{c}\rangle+1)$, equations from Ref. \cite{Kronowetter24} yields 
\begin{align}
\langle\hat{O}_{\rm SMPC}\rangle&=\frac{q_c(1-q_c^K)}{1-q_c},\\
\langle\Delta\hat{O}_{\rm SMPC}^2\rangle&=\frac{q_c}{1-q_c}\left[1-(2K+1)q_c^K+\frac{2q_c(1-q_c^K)}{1-q_c}\right]\nonumber
\\&-\left[\frac{q_c(1-q_c^K)}{1-q_c}\right]^2.
\end{align}
We provide the achievability in the worst resolution case, where $K=1$. Under $K=1$,
\begin{align}
\langle\hat{O}_{\rm SMPC}\rangle&=q_c,\\
\langle\Delta\hat{O}_{\rm SMPC}^2\rangle&=q_c-q_c^2,
\end{align}
or
\begin{align}
\langle\hat{O}_{\rm SMPC}\rangle_{\kappa=0}&=\frac{N_{c,\kappa=0}}{N_{c,\kappa=0}+1},\\
\langle\hat{O}_{\rm SMPC}\rangle_{\kappa>0}&=\frac{N_{c,\kappa>0}}{N_{c,\kappa>0}+1},\\
\langle\Delta\hat{O}_{\rm SMPC}^2\rangle_{\kappa=0}&=\frac{N_{c,\kappa=0}}{(N_{c,\kappa=0}+1)^2},\\
\langle\Delta\hat{O}_{\rm SMPC}^2\rangle_{\kappa>0}&=\frac{N_{c,\kappa>0}}{(N_{c,\kappa>0}+1)^2}.
\end{align}
From $R, N_S\ll 1$, $N_{c,\kappa=0}\simeq N_{c,\kappa>0}\ll1$, and $N_{c,S}\simeq B^2N_B$,
\begin{align}
{\rm SNR_{SMPC}}&=\frac{(\langle\Delta\hat{O}_{\rm SMPC}\rangle_{\kappa>0}-\langle\Delta\hat{O}_{\rm SMPC}\rangle_{\kappa=0})^2}{2\left( \sqrt{\langle\Delta\hat{O}_{\rm SMPC}^2\rangle_{\kappa>0}}+\sqrt{\langle\Delta\hat{O}_{\rm SMPC}^2\rangle_{\kappa=0}}\right)^2}
\\&= \frac{\left(\sqrt{N_{c,\kappa>0}}-\sqrt{N_{c,\kappa=0}}\right)^2}{2\left(1+\sqrt{N_{c,\kappa=0}N_{c,\kappa>0}}\right)^2}
\\&\simeq \frac{\left(\sqrt{N_{c,\kappa>0}}-\sqrt{N_{c,\kappa=0}}\right)^2}{2}
\\&\simeq \frac{\left[(N_{c,\kappa>0}-N_{c,\kappa=0})/\left(2\sqrt{N_{c,\kappa=0}}\right)\right]^2}{2}
\\&=\frac{R(1-R)N_{c,S,I}^2}{2[RN_{c,S}+(1-R)N_{c.I}]}
\\&\simeq \frac{N_{c,S,I}^2}{2N_{c,S}}
\\&\simeq\frac{\eta\kappa N_S}{2N_B}
\end{align}
Hence, $\mathcal{F}_{\rm SMPC}={\rm SNR_{SMPC}}/\gamma_{\rm CI}=2\eta$ is achieved, therefore 3 dB advantage is achievable when $\eta=1$.


\begin{thebibliography}{99}

\bibitem{Shapiro20} J. H. Shapiro, The quantum illumination story, IEEE Aerosp. Electron. Syst. Mag. 35, 8-20 (2020).

\bibitem{Sorelli21} G. Sorelli, N. Treps, F. Grosshans, and F. Boust, Detecting a target with quantum entanglement, IEEE Aerosp. Electron. Syst. Mag. 37, 68 (2021).

\bibitem{Torrome24} R. G. Torromé and S. Barzanjeh, Advances in quantum radar and quantum LiDAR, Progress in Quantum Electronics 93, 100497 (2024).

\bibitem{Karsa23} A. Karsa, A. Fletcher, G. Spedalieri, and S. Pirandola, Quantum Illumination and Quantum Radar: A Brief Overview,  
\textit{arXiv}:2310.06049v1 (2023).

\bibitem{Lloyd08} S. Lloyd, Enhanced sensitivity of photodetection via quantum illumination, Science 321, 1463-5 (2008).

\bibitem{Tan08} S.-H. Tan , B. I. Erkmen, V. Giovannetiti, S. Guha, S. Lloyd, L. Maccone, S. Pirandola, and J. H. Shapiro, Quantum Illumination with Gaussian States, Phys. Rev. Lett. 101, 253601 (2008).

\bibitem{DePalma18} G. De Palma and J. Borregaard, Minimum error probability of quantum illumination, Phys. Rev. A 98, 012101 (2018).

\bibitem{Nair20} R. Nair and M. Gu, Fundamental limits of quantum illumination, Optica 7, 771 (2020). 

\bibitem{Bradshaw21} M. Bradshaw, L. O. Conlon, S. Tserkis, M. Gu, P. K. Lam, and S. M. Assad, Optimal probes for continuous-variable quantum illumination, Phys. Rev. A 103, 062413 (2021).

\bibitem{Zhuang17} Q. Zhuang, Z. Zhang, and J. H. Shapiro, Optimum mixed-state discrimination for noisy entanglement-enhanced sensing, Phys. Rev. Lett. 118, 040801 (2017).

\bibitem{Shi22} H. Shi, B. Zhang, and Z. Zhung, Fulfilling entanglement's optimal advantage via converting correlation to coherence, \textit{arXiv}:2207.-6609 (2022).

\bibitem{Guha09} S. Guha and B. I. Erkmen, Gaussian-state quantum-illumination receivers for target detection, Phys. Rev. A 80, 052310 (2009).

\bibitem{Sanz} M. Sanz, U. Las Heras, J. J. Garcia-Ripoll, E. Solano, and R. Di Candia, Quantum Estimation Methods for Quantum Illumination, Phys. Rev. Lett. 118, 070803 (2017).

\bibitem{Lee} S.-Y. Lee, Y. Jo, T. Jeong, J. Kim, D. H. Kim, D. Kim, D. Y. Kim, Y. S. Ihn, and Z. Kim, Observable bound for Gaussian illumination, Phys. Rev. A 105, 042412 (2022).

\bibitem{Chang19}  C. W. S. Chang, A. M. Vadiraj, and J. Bourassa, Quantum-enhanced noise radar, Appl. Phys. Lett. 114, 112601 (2019).

\bibitem{Luong20} D. Luong, C. W. S. Chang, and A. M. Vadiraj, Receiver operating characteristics for a prototype two-mode squeezing radar, IEEE Aerosp. Electron. Syst. Mag. 56, 2041-2060 (2020).

\bibitem{Barzanjeh20} S. Barzanjeh, S. Pirandola, D. Vitali, and J. M. Fink, Microwave quantum illumination using a digital receiver, Sci. Adv. 6, eabb0451 (2020).

\bibitem{Assouly23} R. Assouly, R. Dassonneville, T. Peronnin, A. Bienfait, and B. Huard, Quantum advantage in microwave quantum radar, Nat. Phys. 19, 1418-1422 (2023).

\bibitem{Zhang15} Z. Zhang, S. Mouradian, F. N. C. Wong, and J. H. Shapiro, Entanglement-Enhanced Sensing in a Lossy and Noisy Environment, Phys. Rev. Lett. 114, 110506 (2015).

\bibitem{England19} D. G. England, B. Balaji, and B. J. Sussman, Quantum-enhanced standoff detection using correlated photon pairs, Phys. Rev. A 99, 023828 (2019).

\bibitem{Aguilar19} G. H. Aguilar, M. A. de Souza, R. M. Gomes, J. Thompson, M. Gu, L. C. Celeri, and S. P. Walborn, Experimental investigation of linear-optics-based quantum target detection, Phys. Rev. A 99, 053813 (2019).

\bibitem{Zhang20} Y. Zhang, D. England, and A. Nomerotski, Multidimensional quantum-enhanced target detection via spectrotemporal-correlation, Phys. Rev. A 101, 053808 (2020).

\bibitem{Xu21} F. Xu, X. M. Zhang, and L. Xu, Experimental Quantum Target Detection Approaching the Fundamental Helstrom Limit, Phys. Rev. Lett. 127, 040504 (2021).

\bibitem{Reichert23} M. Reichert, Q. Zhuang, J. H. Shapiro, and R. Di Candia, Quantum Illumination with a Hetero-Homodyne Receiver and Sequential Detection, Phys. Rev. Applied 20, 014030 (2023)

\bibitem{Kim23} D. H. Kim, Y. Jo, D. Y. Kim, T. Jeong, N. H. Park, Z. Kim, and S.-Y. Lee, Gaussian quantum illumination via monotone metrics, Phys. Rev. Research 5, 033010 (2023).

\bibitem{Lvovsky07} A. I. Lvovsky, B. C. Sanders, and W. Tittel, Optical quantum memory, Nature Photonics 3, 706 (2007).

\bibitem{Tamura17} Y. Tamura, H. Sakuma, K. Morita, M. Suzuki, Y. Yamamoto, K. Shimada, Y. Honma, K. Sohma, T. Fujii, and T. Hasegawa, Lowest-Ever 0.1419-dB/km Loss Optical Fiber, Optical fiber communication Conference, Th5D-1 (2017).

\bibitem{Hisatomi16} R. Hisatomi, A. Osada, Y. Tabuchi, T. Ishikawa, A. Noguchi, R. Yamazaki, K. Usami, and Y. Nakamura, Bidirectional conversion between microwave and light via ferromagnetic magnons, Phys. Rev. B 93, 174427 (2016).

\bibitem{Regal18} A. P. Higginbotham, P. S. Burns, M. D. Urmey, R. W. Peterson, N. S. Kampel, B. M. Brubaker, G. Smith, K. W. Lehnert, and C. A. Regal, Harnessing electro-optic correlations in an efficient mechanical converter, Nat. Phys. 14, 1038 (2018).

\bibitem{Ihn20} Y. S. Ihn, S. Y. Lee, D. Kim, S. H. Yim, and Z. Kim, Coherent multimode conversion from microwave to optical wave via a magnon-cavity hybrid system, Phys. Rev. B 102, 064418 (2020).

\bibitem{Lauk20} N. Lauk, N. Sinclair, S. Barzanjeh, J. P. Covey, M. Saffman, M. Spiropulu, and C. Simon,  Perspective on quantum transduction, Quantum Sci. Technol. 5, 20501 (2020).

\bibitem{Jeong23} T. Jeong, D. H. Kim, D. Kim, Y. S. Ihn, S.-Y. Lee, Y. Jo, J. Kim, Z. Kim, and D. Y. Kim, Polarization-selective optic-to-microwave conversion in a ferromagnetic insulator, J. Appl. Phys. 134, 234402 (2023).

\bibitem{Barzanjeh15} S. Barzanjeh, S. Guha, C. Weedbrook, D. Vitali, J. H. Shapiro and S. Pirandola, Microwave quantum illumination, Phys. Rev. Lett. 114, 080503 (2015).

\bibitem{Ferraro} A. Ferraro, S. Olivares, and M. G. A. Paris, \textit{Gaussian States in Quantum Information} (Bibliopolis, Berkeley, 2005).

\bibitem{Chernoff52} H. Chernoff, A Measure of Asymptotic Efficiency for Tests of a Hypothesis Based on the sum of Observations, Ann. Math. Statist. 23, (4) 493-507 (1952).

\bibitem{Karsa21} A. Karsa and S. Pirandola, Classical benchmarking for microwave quantum illumination, IET Quant. Comm. 2, 246 (2021).



\bibitem{Barzanjeh11} S. Barzanjeh, D. Vitali, P. Tombesi, and G. J. Milburn, Entangling optical and microwave cavity modes by means of a nanomechanical resonator, Phys. Rev. A 84, 042342 (2011).

\bibitem{Ebrahimi22} M. S. Ebrahimi, S. Zippilli, and D. Vitali, Feedback-enabled microwave quantum illumination, Quantum Sci. Technol. 7, 035003 (2022).

\bibitem{Kronowetter24} F. Kronowetter, M. Würth, W. Utschick, R. Gross, K. G. Fedorov, Imperfect photon detection in quantum illumination, Phys. Rev. A 21, 014007 (2024).
\end{thebibliography}
\end{document}